# Suppression of Spin Dephasing in a Two-Dimensional Electron Gas with a Quantum Point Contact

Jae-Seung Jeong and Hyun-Woo Lee

*Department of Physics and PCTP, Pohang University of Science and Technology, Pohang, Gyungbuk, 790-784, Korea*

**Spin-orbit coupling is a source of strong spin dephasing in two- and three-dimensional semiconducting systems. We report that spin dephasing in a two-dimensional electron gas can be suppressed by introducing a quantum point contact. Surprisingly, this suppression was not limited to the vicinity of the contact but extended to the entire two-dimensional electron gas. This facilitates the electrical control of the spin degree of freedom in a two-dimensional electron gas through spin-orbit coupling.**

**Keywords** : spin-orbit coupling, quantum point contact, spin dephasing

## 1. Introduction

One of main aims of semiconductor spintronics [1] is to utilize the spin-orbit coupling to control the electron spin degree of freedom [2-4]. In this respect, the Rashba spin-orbit (RSO) coupling [5] arising in a two-dimensional electron gas (2DEG) plays an important role because its strength can be modulated electrically [6]. This opens the possibility of achieving electrical control of the spin degree of freedom.

The RSO coupling controls the spin by generating an effective magnetic field, around which the spins precess. However, unlike conventional magnetic fields, this effective magnetic field varies with the momentum of electrons, which causes the electron spins to precess around different axes depending on their momenta. The angle average over the momentum direction then results in spin dephasing [7]. This RSO coupling-induced spin dephasing is strong even in a ballistic 2DEG [8, 9]. Therefore, it is important to suppress this adverse effect of RSO coupling to achieve efficient electrical control of the spin degree of freedom.

A narrow 2DEG with only one transport channel is an ideal environment for spin transport because the RSO coupling-induced spin dephasing is quenched [2, 10]. However, such a single-channel system is rather difficult to realize in experiments. For example, the width of the system needs to be smaller than 10 nm for a 2DEG with a Fermi wavelength of 10 nm [8]. The other way to produce a single-channel system is to introduce a quantum point contact (QPC) in a 2DEG. It is well known that the number of transport channels decreases via a QPC when charge conductance is observed [11]. For spintronic applications, the injection and detection of the spin-polarized current was achieved experimentally using a QPC and external magnetic field [12]. Furthermore, experimental evidence suggests that the spin-orbit coupling caused by the highly asymmetric lateral potential in the QPC can generate a spin-polarized current [13]. However, there are no reports of the RSO coupling-induced spin precession of the spin-polarized current through the QPC, which would be due to the decrease in the number of channels.

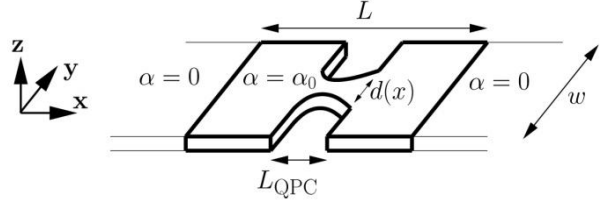

Fig. 1. Schematic diagram of a two-dimensional electron gas with a quantum point contact.

This paper reports that the RSO coupling-induced spin dephasing can be strongly suppressed in a 2DEG with a QPC (Fig. 1). Transport through the QPC becomes one-dimensional (1D) when its channel width $d(x)$ is comparable to the Fermi wave length $\lambda_F$ in the 2DEG. A trivial effect of the QPC is the suppression of the spin dephasing in the vicinity of the QPC. This is because in a 1D wire, the electron motion perpendicular to the wire axis is strongly quenched, preventing the angle average over the momentum direction. In addition to this trivial effect, the QPC causes a far reaching nonlocal effect; the spin dephasing is suppressed not only in the vicinity of the QPC but also in the entire region of the 2DEG. This paper demonstrates this in the ballistic regime (free from impurities) and then in the weakly diffusive regime. It is believed that this method will facilitate efficient electrical control of the spin degree of freedom in a 2DEG via the RSO coupling.

## 2. Theoretical Model

We first consider a ballistic 2DEG subject to the RSO coupling, which is described (Fig. 1) by the Hamiltonian $H$,

$$H = \frac{p_x^2 + p_y^2}{2m^*} + \frac{\{\alpha(x),\ \sigma_x p_y - \sigma_y p_x\}}{2\hbar} + V_c(y) + V_{QPC}(x, y) \quad (1)$$

where $m^*$ is the effective mass of an electron, $\boldsymbol{\sigma} = (\sigma_x, \sigma_y, \sigma_z)$ is the Pauli spin operator. Here, the first, second, third and fourth terms represent respectively, the kinetic energy, the RSO coupling, the hard wall potential at the side edges ($y = \pm w/2$) of the 2DEG, and hard wall potential that generates the QPC in the range $|x| < L_{QPC}/2$, whose width is $d(x)$ [14].

To illustrate the main physics, the RSO coupling strength $\alpha(x)$ was chosen to have a nonzero value $\alpha_0$ only within a finite range ($|x| < L/2$) and disappear elsewhere [15]. Hence, the spin becomes a conserved quantum number in the left ($x < -L/2$) and right ($x > L/2$) ``electrodes''. The degree of spin dephasing caused within the range $|x| < L/2$ can then be obtained from the spin-resolved conductance from one electrode to the other.

## 3. Results and Discussion

Figure 2 (a) shows the normalized spin-resolved conductance $g^{+\mathbf{x},+\mathbf{x}}(\alpha_0) \equiv G^{+\mathbf{x},+\mathbf{x}}(\alpha_0) / G^{+\mathbf{x},+\mathbf{x}}(\alpha_0=0)$ as a function of $\alpha_0$, where $g^{+\mathbf{x},+\mathbf{x}}(\alpha_0) = (e^2/h) \sum_{i,j} |t_{i,j}^{+\mathbf{x},+\mathbf{x}}|^2$ was calculated within the Landauer-Büttiker formalism [16, 17]. Here $t_{i,j}^{+\mathbf{n},+\mathbf{n}}$ denotes the transmission amplitude from the orbital channel $j$ on the left electrode with spin pointing in the $+\mathbf{n}$-direction to the orbital channel $i$ on the right electrode with spin pointing in the $+\mathbf{n}$-direction. The normalized conductance $g^{+\mathbf{x},+\mathbf{x}}(\alpha_0)$ amounts to the probability that an electron, whose spin is initially aligned along the $+\mathbf{x}$-direction in the left electrode, arrives the right electrode with its spin aligned along the $+\mathbf{x}$-direction. The sinusoidal oscillation of $g^{+\mathbf{x},+\mathbf{x}}(\alpha_0)$ in Fig. 2 (a) is due to the spin precession caused by the RSO coupling. In the absence of the QPC ($V_{QPC}=0$ and $d(0)=w$), the oscillation amplitude decays fast with increasing $\alpha_0$. Following the abnormal behavior near $\alpha_0 \sim 10 \times 10^{-12}$ eVm, which is a trace of the beating phenomenon [18], the $g^{+\mathbf{x},+\mathbf{x}}(\alpha_0)$ at both peaks and dips converge towards 0.5, representing a strong spin dephasing effect of the RSO coupling.

Figure 2 (a) also shows the results in the presence of the QPC. In Fig.2 (a), $N_{QPC} \equiv \text{Int}[d(0)/(\lambda_F/2)]$ is a measure of the width of the QPC and represents the upper bound of $g^{+\mathbf{x},+\mathbf{x}}(\alpha_0)/(e^2/h)$ imposed by the conductance quantization effect of the QPC [17, 19]. Here, Int[$q$] represents the largest integer not exceeding $q$. It should be noted that the QPC enhances the oscillation amplitude of $g^{+\mathbf{x},+\mathbf{x}}(\alpha_0)$ and the enhancement increases with decreasing $N_{QPC}$.

As a measure of the spin coherence, we introduce the spin-orbit resistance (SOR), which is defined as

$$\text{SOR} \equiv \frac{G_{max} - G_{min}}{G_{min}} \times 100\% \tag{2}$$

where $G_{max}$ and $G_{min}$ represent the local maximum and minimum values of $G^{+\mathbf{x},+\mathbf{x}}(\alpha_0)$, respectively. Figure 2 (b) and (c) were obtained by evaluating the maximum and minimum in the range $4 \times 10^{-12}$ eVm $< \alpha_0 < 9 \times 10^{-12}$ eVm [6, 20] and by evaluating them in the range $45 \times 10^{-12}$ eVm $< \alpha_0 < 50 \times 10^{-12}$ eVm, respectively [21]. Many experiments [6, 20, 21] reported $\alpha_0$ in these ranges. Both in Figs. 2 (b) and (c), the SOR was plotted as a function of $w/(\lambda_F/2)$, which is a measure of the width of the 2DEG and whose integer part Int[$w/(\lambda_F/2)$] represents the upper bound [17] of $G^{+\mathbf{x},+\mathbf{x}}(\alpha_0)/(e^2/h)$ imposed by the finite width $w$ of the 2DEG. In the absence of the QPC, the SOR decays rapidly with increasing $w/(\lambda_F/2)$. On the other hand, the SOR does not exhibit such decay in the presence of the QPC. Instead it appears to saturate after initial fluctuations that originated from relative large variations in the QPC shape as $w/(\lambda_F/2)$ changes in its smaller ranges. This demonstrates that spin dephasing can be suppressed considerably by the QPC.

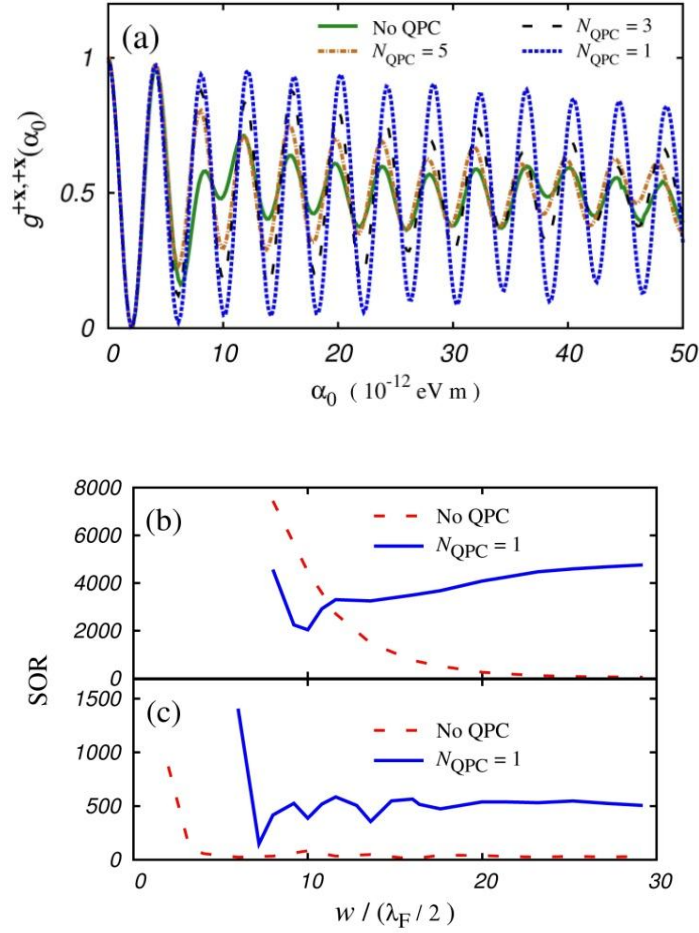

Fig. 2. (Color online) (a) The normalized spin-resolved conductance $g^{+\mathbf{x},+\mathbf{x}}(\alpha_0) \equiv G^{+\mathbf{x},+\mathbf{x}}(\alpha_0) / G^{+\mathbf{x},+\mathbf{x}}(\alpha_0=0)$ in the ballistic 2DEG as a function of $\alpha_0$ in the presence or absence of the QPC. $w/(\lambda_F/2) = 20.1$ in this plot. (b,c) The SOR [Eq. (2)] as a function of $w/(\lambda_F/2)$ in the presence ($N_{QPC}=1$) or absence of the QPC. The SOR was evaluated in the range $4\times10^{-12}$ eVm $< \alpha_0 < 9\times10^{-12}$ eVm [6, 20] for (b) and $45\times10^{-12}$ eVm $< \alpha_0 < 50\times10^{-12}$ eVm [21] for (c), respectively. In the numerical calculations, $m^* = 0.04 m_{electron}$, $\lambda_F = 193$A, $L=176\,\lambda_F$, $L_{QPC} = 17\,\lambda_F$. Here $m_{electron}$ is the free electron mass.

The following discusses the origin of the spin dephasing suppression. Semiclassicaly, the total spin dephasing probability is the sum of the spin dephasing probabilities in the QPC region ($|x|<L_{QPC}/2$) and outside regions ($L_{QPC}/2 < |x| < L/2$). In this semiclassical picture, the QPC cannot enhance the spin coherence by more than 22% because $L_{QPC}/L=0.22$. In contrast, Figs 2 (b) and (c) show much more significant enhancement of the spin coherence. Therefore, this semiclassical picture fails.

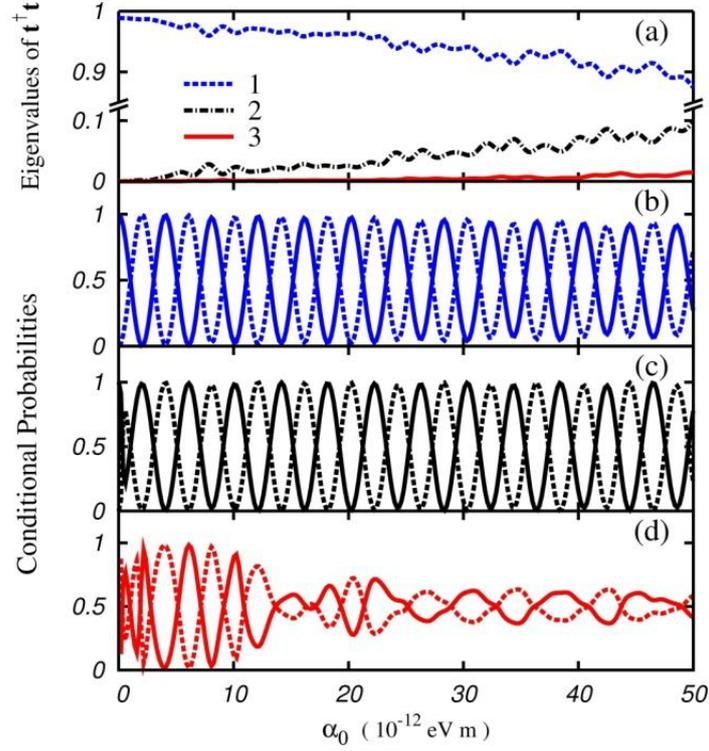

Fig. 3. (Color online) (a) The transmission probabilities (eigenvalues of $\mathbf{t}^\dagger\mathbf{t}$) of the eigen-transport channels 1, 2, and 3 for $N_{QPC}=1$ and $w/(\lambda_F/2) = 20.1$. Other eigen-transport channels (not shown) have smaller transmission probabilities. The conditional probabilities of the eigen-transport channels 1 (b), 2 (c), and 3 (d) have spin $+\mathbf{x}$ (solid lines) or $-\mathbf{x}$ (dashed lines). See the text for more details.

A nontrivial quantum effect of the QPC was recently demonstrated by Eto *et al*. [19]. It is well known that the RSO coupling may induce the anti-crossing between energy bands with different spin directions. They demonstrated that the anti-crossing near the QPC might result in spin filtering. However, this spin filtering effect does not explain the result in Fig. 2. To begin with, this mechanism can only function when the anti-crossing of the local energy bands occurs near the QPC. For $N_{QPC}=1$, this requires $m^*\alpha_0/\hbar^2 k_F$ to be larger than $\sim 0.1$ according to the estimation in Ref. [19]. This requirement is satisfied only for sufficiently large $\alpha_0 > 30\times 10^{-12}$ eVm while the SOR enhancement by the QPC occurs for a much smaller $\alpha_0$. Secondly and more importantly, this spin filtering mechanism polarizes the spin along the eigen-spin direction of the spin-orbit coupling, which is along the $\pm\mathbf{y}$-direction. Therefore, this mechanism actually suppresses the spin precession within the xz-plane and reduces the oscillation amplitude of $g^{+\mathbf{x},+\mathbf{x}}(\alpha_0)$. For these two reasons, it is concluded that this mechanism cannot explain the result in Fig. 2 [22].

The behaviors of eigen-transport channels were examined to gain insight into the origin of the spin dephasing suppression [11]. The concept of the eigen-transport channels is a very successful tool for understanding various mesoscopic phenomena, such as the conductance quantization [17, 23], universal conductance fluctuations [24] and shot noise [25]. For the structure in Fig. 1, they are defined as eigenvectors of the matrix $\mathbf{t}^\dagger \mathbf{t}$, where the matrix elements of $\mathbf{t}$ are given by $\mathbf{t}_{i,j}^{s,+\mathbf{x}}$ ($s = +\mathbf{x}, -\mathbf{x}$) and contain all information about the electrons injected with $+\mathbf{x}$ spin. For $\alpha_0 = 0$, the QPC introduces a small number of eigen-transport channels with transmission probabilities (eigenvalues of $\mathbf{t}^\dagger \mathbf{t}$) close to one, while the remaining eigen-transport channels have much smaller transmission probabilities close to zero [11].

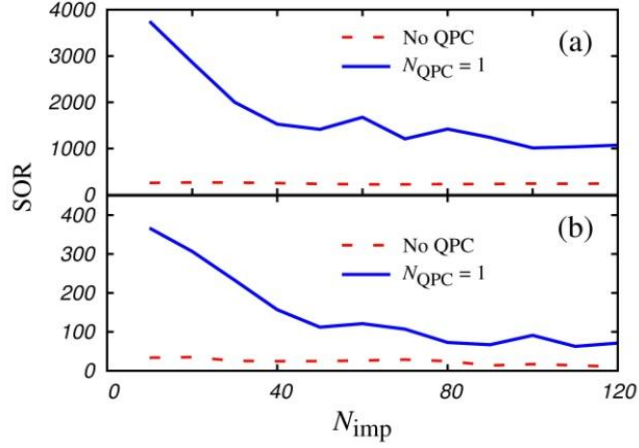

Fig. 4. (Color online) The SOR [Eq. (2)] as a function of the number of scatterers $N_{\text{imp}}$ for a system with $w/(\lambda_F/2) = 20.1$. The SOR in (a) and (b) are evaluated in the $\alpha_0$ ranges used in Figs. 2 (b) and (c), respectively.

Figure 3 (a) shows the three largest eigenvalues of $\mathbf{t}^\dagger \mathbf{t}$ as a function of $\alpha_0$ for $N_{\text{QPC}}=1$. Although the separation between eigen-transport channels with transmission probabilities close to one and zero becomes less clear with increasing $\alpha_0$, one particular eigen-transport channel (channel 1) is still dominant over the others. Figure 3 (b), (c) and (d) show the conditional probabilities that an electron has its spin pointing in the $+\mathbf{x}$ (solid lines) and $-\mathbf{x}$-directions (dashed lines) when it arrives at the right electrode. Note that the conditional probability of eigen-transport channel 3 shows irregular oscillations for $\alpha_0 > 13 \times 10^{-12}$ eVm, indicating the strong influence of the spin dephasing in the 2DEG. Interestingly, all such eigen-transport channels with irregular oscillations have small transmission probabilities. In contrast, eigen-transport channels 1 and 2, which are the best and second best transmission channels, show regular oscillations with large amplitudes. However, to be more precise, the oscillation of the eigen-transport channel 2 is not strictly regular either. A closer look shows an incomplete oscillation near $\alpha_0 = 1.2 \times 10^{-12}$ eVm. Because of this, the conditional probability oscillation of eigen-transport channel 2 is 180° out of phase from that of eigen-transport channel 1 and reduces the oscillation amplitude of $g^{+\mathbf{x},+\mathbf{x}}(\alpha_0)$. The degree of the reduction depends on the relative magnitude of the transmission probabilities of the two eigen-transport channels. Moreover, the oscillation amplitude of $g^{+\mathbf{x},+\mathbf{x}}(\alpha_0)$ can remain large because the transmission probability for eigen-transport channel 1 is much larger than that for eigen-transport channel 2. Therefore, the QPC allows a good transmission for an eigen-transport channel, which is the least affected by the spin dephasing, and suppresses the transmissions from the eigen-transport channels affected significantly by the spin dephasing. This selective transmission obviously suppresses the spin dephasing as well as the entanglement between the spin and orbital degrees of freedom, which is an important source of the spin dephasing [8]. Recalling that the electron motion near the QPC is 1D-like, the best transmission channel also has strong 1D-character [Fig. 3 (b)]. In addition, this can explain why the spin dephasing suppression is not limited to the vicinity of the QPC because the transmission amplitude **t** depends on the structure of the entire system.

The effects of the QPC in the weakly diffusive regime were also examined. In the numerical conductance calculation, strongly repulsive scatterers are introduced at $N_{imp}$ randomly selected sites in the 2DEG ($L_{QPC}/2 < |x| < L/2$) to study the scattering effects. Figure 4 shows the SOR as a function of $N_{imp}$ for a system with the QPC ($N_{QPC}=1$) and $w/(\lambda_F/2) = 20.1$. Although the SOR decreases with $N_{imp}$, it remains, up to $N_{imp} \sim 30$, more than one order of magnitude higher than the corresponding value in the absence of the QPC. Moreover, even for $N_{imp} \sim 120$, it still remains significantly larger than the corresponding value in the absence of the QPC. For $N_{imp} = 120$, the mean free path $l$ is approximately one third of $L$ [26]. Therefore, even in the weakly diffusive regime, the presence of the QPC enhances the SOR considerably. As a passing remark, it should be noted that each data point in Fig. 4 was obtained for one particular random configuration of scatterers. To check if the results are generic, a few other configurations were tested and qualitatively similar results were obtained.

Previously it was reported [3, 27] that when an electron is subject to both the RSO coupling and the Dresselhaus spin-orbit (DSO) coupling [28], the spin dephasing can be suppressed by exploiting the competition of the two types of the spin-orbit coupling. Although this suppression mechanism is robust against diffusive scattering, it works only at a special value of $\alpha_0$ [3, 27]. In contrast, the suppression by the QPC works over a wide range of $\alpha_0$ provided $l/L \geq 0.3$. In this sense, these two mechanisms are complementary. In addition, the suppression by the QPC is not sensitive to the detailed forms of the spin-orbit coupling. As a test, it was confirmed that similar suppression persists when the RSO coupling coexists with the linear DSO coupling for the crystal direction [110].

The introduction of the QPC may cause some spatial changes in $\alpha_0$ near the QPC, and the gate voltage, which is intended to control $\alpha_0$, might also affect the width $d(x)$ of the QPC. However these complications do not deteriorate the spin dephasing suppression significantly provided the spatial variation of $\alpha_0$ does not cause significant reflection of electrons [15] and the change in $d(x)$ does not alter $N_{QPC}$.

## 4. Summary

The introduction of a QPC can strongly suppress the spin dephasing in a 2DEG. The orders of magnitude enhancement of the SOR [Eq. (2)] can be obtained using this method. It is believed that this result will facilitate the electrical control of the spin degree of freedom in a 2DEG, and when combined with recent progresses in spin injection and detection [29] it can make the realization of the spin transistor [2] feasible

## Acknowledgement

This study is supported by the KOSEF (R11-2000-071, R01-2007-000-20281-0) and the KRF (KRF-2005-070-C00055). J.-S.J thanks Dong-In Chang for valuable discussion about experimental conditions. The authors wish to thank Beom Hyun Kim, Hyowon Park, and Heung-Sun Sim for their help on the numerical calculations. We also acknowledge the hospitality of Prof. Hyunsoo Yang. Parts of this work were performed while visiting his lab.

## References

[1] I. Žutic, J. Fabian and S. Das Sarma, Rev. Mod. Phys. **76**, 323 (2004); D. D. Awschalom and M. E. Flatte, Nature Phys. **3**, 153 (2007).

[2] S. Datta and B. Das, Appl. Phys. Lett. **56**, 665 (1990); H.-C. Koo, J.-H. Kwon, J. Eom, J. Chang, S. H. Han, and M. Johnson, Science **325**, 1515 (2009).

[3] J. Schliemann, J. C. Egues, and D. Loss, Phys. Rev. Lett. **90**, 146801 (2003).

[4] F. Zhai and H. Q. Xu, Phys. Rev. Lett. **94**, 246601 (2005).

[5] Yu. A. Bychkov and E. I. Rashba, J. Phys. C **17**, 6039 (1984); JETP Lett. **39**, 78 (1984).

[6] J. Nitta, T. Akazaki, and T. Enoki, Phys. Rev. Lett. **78**, 1335 (1997); T. Koga, J. Nitta, T. Akazaki, and H. Takayanagi, *ibid.* **89**, 046801 (2002).

[7] M. I. D'yakonov and V. I. Perel', Fiz. Tverd. Tela (Leningrad) **13**, 3581 (1971) [Sov. Phys. Solid State **13**, 3023 (1971)].

[8] B. K. Nikolić and S. Souma, Phys. Rev. B **71**, 195328 (2005).

[9] J.-S. Jeong and H.-W. Lee, Phys. Rev. B **74**, 195311 (2006).

[10] S. Bandyopadhyay and M. Cahey, Appl. Phys. Lett. **85**, 1433 (2004); S. Pramanik, S. Bandyopadhyay, and M. Cahay, IEEE Trans. Nanotechnol. **4**, 2 (2005).

[11] A. D. Stone, P. A. Mello, K. A. Muttalib, and J.-L. Pichard, in *Mesoscopic Phenomena in Solids*, edited by B. L. Al'tshuler, P. A. Lee, and R. A. Webb (North-Holland, Amsterdam, 1991).

[12] R. M. Potok, J. A. Folk, C. M. Marcus, and V. Umansky, Phys. Rev. Lett. **89**, 266602 (2002); E. J. Koop, B. J. van Wees, D. Reuter, A. D. Wieck, and C. H. van der Wal, *ibid*. **101**, 056602 (2008); S. M. Frolov, A. Venkatesan, W. Yu, J. A. Folk, and W. Wegscheider, *ibid*. **102**, 115802 (2009).

[13] P. Debray, S. M. S. Rahman, J. Wan, R. S. Newrock, M. Cahay, A. T. Ngo, S. E. Ulloa, S. T. Herbert, M. Muhammad, and M. Johnson, Nature Nanotechnology **4**, 759 (2009).

[14] In numerical calculation, we use $d(x) = d_0 + 2S_{QPC} |x|$ with $S_{QPC}$ 0.25. For $N_{QPC} = 1$, we use $d_0 = 0.60\lambda_F$. $d_0$ increases $N_{QPC}$.

[15] The electron reflection probability due to the abrupt change of $\alpha(x)$ at $|x| = L/2$ is negligible when the RSO coupling energy $m^*\alpha_0^2/\hbar^2$ is sufficiently smaller than the Fermi energy, which is indeed the case in most experimental situations.

[16] M. P. Anantram and T. R. Govindan, Phys. Rev. B **58**, 4882 (1998).

[17] S. Datta, *Electronic transport in mesoscopic systems* (Cambridge University Press, Cambridge, 1995).

[18] X. F. Wang, Phys. Rev. B **69**, 035302 (2004); L. Zhang, P. Brusheim, and H. Q. Xu, *ibid.* **72**, 045347 (2005).

[19] M. Eto, T. Hayashi, and Y. Kurotani, J. Phys. Soc. Jpn. **74**, 1934 (2005).

[20] J. Luo, H. Munekata, F. F. Fang, and P. J. Stiles, Phys. Rev. B **41**, 7685 (1990); G. Engels, J. Lange, Th. Schäpers, and H. Lűth, Phys. Rev. B **55**, R1958 (1997); J. P. Heida, B. J. van Wees, J. J. Kuipers, T. M. Klapwijk, and G. Borghs, Phys. Rev. B **57**, 11911 (1998).

[21] M. Schultz, F. Heinrichs, U. Merkt, T. Colin, T. Skauli, and S. Lovold, Semicond. Sci. Technol. **11**, 1168 (1996); J. Hong, J. Lee, S. Joo, K. Rhie, B. C. Lee, J. Lee, S. An, J. Kim, and K.-H. Shin, J. Korean Phys. Soc. **45**, 197 (2004).

[22] The enhanced electron-electron interaction near a QPC [K. J. Thomas, J. T. Nicholls, M. Y. Simmens, M. Pepper, D. R. Mace, and D. A. Ritchie, Phys. Rev. Lett. **77**, 135 (1996)] may affect the spin dephasing. This effect however goes beyond the scope of the present paper.

[23] L. I. Glazman, G. B. Lesovik, D. E. Khmel'nitskii, and R. I. Shekhter, Pis'ma Zh. Eksp. Teor. Fiz. **48**, 218 (1988) [JETP Lett. **48**, 238 (1988)]; A. Szafer and A. D. Stone, Phys. Rev. Lett. **62**, 300 (1989).

[24] Y. Imry, Europhys. Lett. **1**, 249 (1986).

[25] G. B. Lesovik, JETP Lett. **49**, 592 (1989); S. -R. E. Yang, Solid State Commun. **81**, 375 (1992).

[26] $G^{+\mathbf{x},+\mathbf{x}}(\alpha_0)$ in the absence of the QPC is evaluated in two different situations; $N_{imp} = 120$ and $N_{imp} = 0$. Their ratio then provides an estimation of $l/L$.

[27] A. A. Kiselev and K. W. Kim, Phys. Status. Solidi (b) **221**, 491 (2000).

[28] G. Dresselhaus, Phys. Rev. **100**, 580 (1955).

[29] S. A. Crooker, M. Furis, X. Lou, C. Adelmann, D. L. Smith, C. J. Palmstrøm, and P. A. Crowell, Science **309**, 2191 (2005); B.-C. Min, K. Motohashi, C. Lodder, and R. Jansen, Nature Mater. **5**, 817 (2006); X. Lou, C. Adelmann, S. Crooker, E. S. Garlid, J. Zhang, K. S. M. Reddy, S. D. Flexner, C. J. Palmstrøm, and P. A. Crowell, Nature Phys. **3**, 197 (2007); H. C. Koo, H. Yi, J.-B. Ko, J. Chang, S.-H. Han, D. Jung, S.-G. Huh, and J. Eom, Appl. Phys. Lett. **90**, 022101 (2007); M. Ramsteiner, O. Brandt, T. Flissikowski, H. T. Grahn, M. Hashimoto, J. Herfort, and H. Kostial, Phys. Rev. B **78**, 121303(R) (2008).